\documentclass{ws-procs10x7}

\newcolumntype{d}[1]{D{.}{.}{#1}}

\newcommand {\Dzero} {D\O\ }

\newcommand{\MET}{\mbox{$E\kern-0.50em\raise0.10ex\hbox{/}_{T}$}}
\newcommand{\vecMET}{\mbox{$\vec{E}\kern-0.50em\raise0.10ex\hbox{/}_{T}$}}

\newcommand{\ttbar}{t \overline{t}}
\newcommand{\ppbar}{p \overline{p}}

\newcommand{\epem}{ e^+e^-}
\newcommand{\pt}{p_{T}}

\makeindex
\begin{document}

\title{$WW$ and $WZ$ Production at the Tevatron}

\author{E. Lipeles}

\address{University of California, San Diego \\E-mail: lipeles@fnal.gov}


\twocolumn[\maketitle\abstract{This report summarizes recent measurements of the
production properties of $WW$ and $WZ$ pairs of bosons at the Tevatron. This includes
measurements of the cross-section and triple gauge couplings in the $WW$ process
and the first evidence for $WZ$ production.}
]

\section{Introduction}
\label{sec:intro}

Boson pair production is one of the few processes that have significant
effects from triple boson vertices at tree level. These couplings are predicted 
in the standard model and are directly related to its gauge group structure. One
of the goals of diboson measurements is to limit deviations from the standard model
values of these triple gauge couplings (TGCs). Such deviations could be observed in 
either the cross-sections or in the kinematic distributions of the observed events.
Possible causes of anomalous TGCs include new particles in loop diagrams.\cite{ref:threv}
It is also possible for diboson final states to receive contributions from 
the $s$-channel production of an as yet unobserved
particle, most notably the standard model Higgs decaying to a pair of $W$ bosons.

\begin{figure}
\begin{center}
\unitlength=0.50\textwidth
\includegraphics[width=0.38\textwidth]{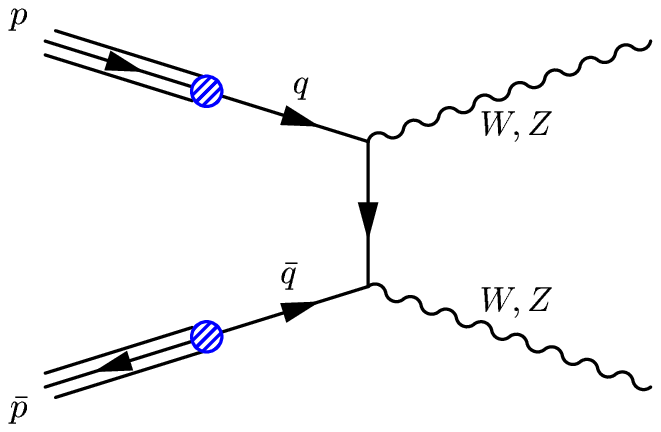}
\put(-0.83,0.4){(a)} 
\vspace{0.5cm}
\includegraphics[width=0.38\textwidth]{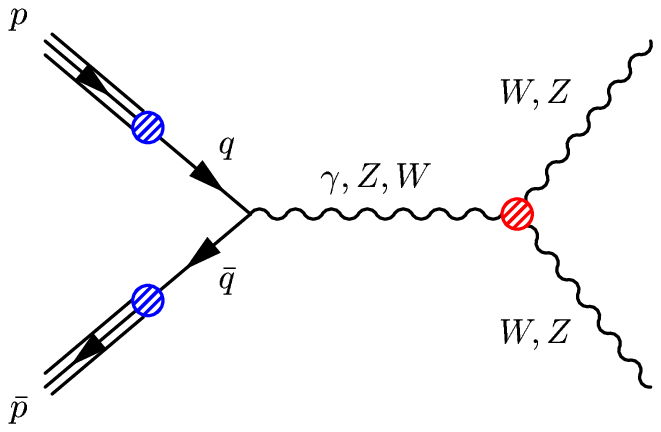}
\put(-0.83,0.4){(b)}
\end{center}
\caption{Dominant diagrams in the production of boson pairs: (a) $t$-channel (b) $s$-channel.}
\label{fig:feyndiags}
\end{figure}

This report summarizes recent measurements by the CDF and \Dzero collaborations
of $WW$ and $WZ$ production at the Tevatron. The Tevatron produces $\ppbar$ collision
at 1.96 TeV center of mass energy. The dominant contributions to the cross-sections for
$WW$ and $WZ$ production are the $t$-channel (and similar $u$-channel) process involving 
two instances of the well measured boson-quark couplings and the $s$-channel process involving 
triple gauge
couplings, shown in Figure \ref{fig:feyndiags}.
The TGCs can in general be functions of the invariant mass of the final state bosons $\sqrt{\hat{s}}$,
so measurements at the Tevatron compliment previous measurements at LEP because they probe
larger values of $\sqrt{\hat{s}}$. Furthermore, the $WZ$ final state, which is not accessible in $\epem$ 
collisions, isolates the $WWZ$ coupling from $WW\gamma$.

Of the heavy diboson processes, $WW$ production is the largest with a standard model 
next-to-leading order (NLO) prediction of $\sigma(WW)_{NLO} = 12.4 \pm 0.8~\rm{pb}$ , followed 
by $WZ$ production with an NLO prediction 
of $\sigma(WZ)_{NLO} = 3.7 \pm 0.1~\rm{pb}$.\cite{ref:crosssections_th}
Section \ref{sec:WWxsec} of this report covers measurements of the $WW$ cross-section;
Section \ref{sec:AnomCoup} describes measurements of the triple gauge couplings in $WW$ production,
and Section \ref{sec:WZ} presents the first evidence for $WZ$ production. 

\section{$WW$ cross-section}
\label{sec:WWxsec}

Both CDF and \Dzero have measured the $WW$ production cross-section in the 
all leptonic final state $l\nu l'\nu$, where $l,l'=e$ or $\mu$. Although this 
is the lowest branching fraction final state of $WW$ decay (4.6\%), it has 
significantly lower backgrounds than the other final states, which all involve
hadronic jets. The presence of neutrinos is identified as lack momentum 
balance in the plane transverse to the beam
using the missing transverse energy variable $\vecMET \equiv \sum_i E_i \hat{n}_T^i$,
where $\hat{n}_T^i$ is the transverse component of a unit vector connecting 
the interaction point to a calorimeter cell $i$ and  $E_i$ is the energy deposition
in that cell. \Dzero requires $\MET\equiv|\vecMET|>30 (ee), 40 (\mu\mu), 20 (e\mu)$ GeV while
CDF requires $\MET> 25$ GeV for all final states.

The primary backgrounds to the $ll'\MET$ final state are from $W\rightarrow l \nu$
with an associated jet or photon which is misidentified as another lepton,
Drell-Yan ($Z/\gamma^*$) production of lepton pairs combined with large false 
$\MET$ due to detector effects, $\ttbar\rightarrow WWbb \rightarrow ll'\nu\nu bb$, 
and other heavy dibosons, either $WZ\rightarrow lll\nu$, with a lost lepton, or 
$ZZ\rightarrow l l \nu\nu$. The predicted composition of the selected sample
in the CDF analysis is shown in Table \ref{tab:WWyields}.

\begin{table}
\tbl{Expected composition and observed yield for the CDF $WW$ analysis. \label{tab:WWyields}}
{\begin{tabular}{@{}lr@{}}
\toprule
Mode                                     &            Events $\pm$ Stat $\pm$ Syst \\
\colrule
Drell-Yan                                &       11.8 $\pm$ 0.8 $\pm$ 3.1    \\
$W$+jets                                 &       11.0 $\pm$ 0.5 $\pm$ 3.2    \\
$WZ+ZZ$                                  &        7.9 $\pm$ 0.0 $\pm$ 0.8    \\
$W\gamma$                                &        6.8 $\pm$ 0.2 $\pm$ 1.4    \\
$\ttbar$                                 &        0.2 $\pm$ 0.0 $\pm$ 0.0    \\
\colrule
Sum Bkg                                  &       37.8 $\pm$ 0.9 $\pm$ 4.7   \\
$WW$                                     &       52.4 $\pm$ 0.1 $\pm$ 4.3    \\
\colrule
Expected                                 &      90.2$\pm$ 0.9 $\pm$ 6.4   \\
\colrule
Data                                     &         95   \\
\botrule
\end{tabular}}
\end{table}

After selection of events with two leptons, significant $\MET$ and a jet veto
to suppress the $\ttbar$ background, CDF observes 95 events in $825~\rm{pb}^{-1}$ of data
with an expected background of 37.8$\pm$0.9(stat.)$\pm$4.7(syst.)\cite{ref:CDFWW} 
and \Dzero observes 25 events in $224-252~\rm{pb}^{-1}$ of data with an expected background of 
8.1$\pm$0.6(stat)$\pm$ 0.6(sys)$\pm$0.5(lumi).\cite{ref:D0WW} These observations correspond to 
cross-sections of $\sigma(WW) = 13.6 \pm 2.3 (stat) \pm 1.6 (sys) \pm 1.2 (lum) ~\rm{pb}$ (CDF)
and $\sigma(WW) = 13.8 ^{+4.3}_{-3.8}(stat) ^{+1.2}_{-0.9}(sys) \pm 0.9 (lum) ~\rm{pb}$ (\Dzero),
both of which are consistent with a NLO calculation of the standard model expectation.

\section{Triple gauge couplings in $WW$ and $WZ$}
\label{sec:AnomCoup}

\begin{figure*}[!thb]
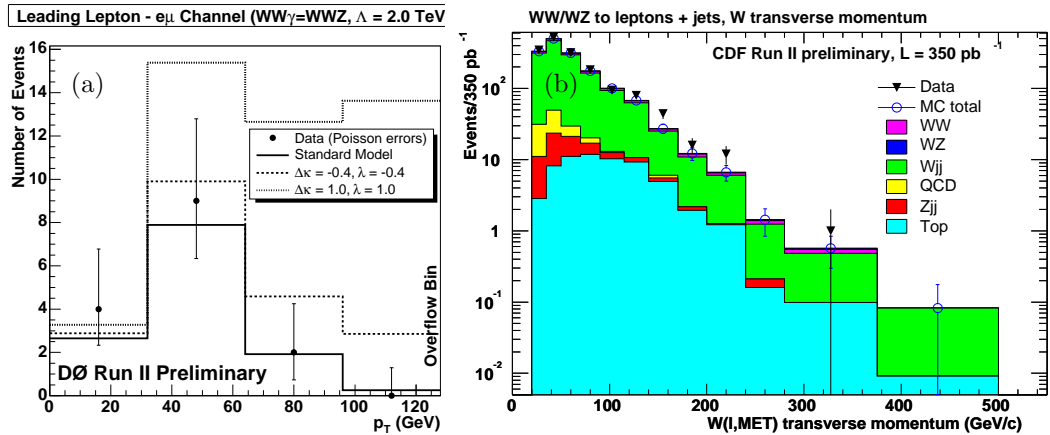

\unitlength=0.50\textwidth
\includegraphics[width=0.43\textwidth]{D0_WW_PT_AnonCoup.epsi}
\put(-0.74,0.65){(a)} 
\includegraphics[width=0.56\textwidth]{CDF_wwwz_jj_wpt_dist.epsi}
\put(-0.96,0.65){(b)}
\caption{Distributions which are fit for triple gauge couplings: 
(a) leading lepton $\pt$ in the $ll'\nu\nu$ final state at \Dzero and
(b) leptonically decaying $W$ $\pt$  in the $l\nu jj$ final state at CDF.}
\label{fig:TGC_fit}
\end{figure*}

The $s$-channel $WW$ production process (shown in Figure \ref{fig:feyndiags}b)
has contributions from both $WW\gamma$ and $WWZ$ TGCs while the $WZ$ process
only gets a contribution from the $WWZ$ vertex. \Dzero has set limits on anomalous couplings
in the $WW\rightarrow ll'\nu\nu$ sample described above.

CDF has recently probed
the TGCs in the combination of $WW$ and $WZ$ modes using the $l\nu jj$
final state, where $l=e$ or $\mu$. While the background in the $l\nu jj$ is
dramatically larger than the purely leptonic final states due to the 
$\ppbar\rightarrow W$+jets process, the branching fractions are $\approx$6.5 times 
higher for $WW$ and $\approx$10 times higher for $WZ$. Because anomalous TGCs
are expected to enhance the high $W$ transverse momentum region, where the backgrounds
from $W$+jets are smaller, the $l\nu jj$ final state can be sensitive to anomalous
couplings even without observation of the standard model processes. 

The TGCs are parameterized by adding terms with variable coupling constants
to the standard model Lagrangian
\begin{eqnarray*}
{\cal L}_{WWV} &=&
 i { {g_1^V} (W_{\mu\nu} W^\mu V^\nu -W_\mu V_\nu W^{\mu\nu})}  \cr 
&+& i { {\kappa_V} W_{\mu} W_\nu V^{\mu\nu}} + 
 \frac{i { {\lambda_{V}}}}{{ M_W^2}} { W_{\lambda\mu} W_\nu^\mu
V^{\nu\lambda}},
\end{eqnarray*}
where ${\kappa_\gamma} = {\kappa_{Z}} = {g_1^Z} = 1$ and ${\lambda_{Z}} = {\lambda_{\gamma}} = 0$
in the standard model. These couplings can be related to the electric
and magnetic dipole and quadrupole moments of the $W$ and $Z$.  In general 
the coupling constants 
$\alpha={\kappa_\gamma},{\kappa_{Z}},{g_1^Z},{\lambda_{Z}},$ and 
${\lambda_{\gamma}}$ can be functions of the invariant mass of the diboson pair $\sqrt{\hat{s}}$.
As a simplification, both experiments assume the functional form 
$\alpha(\hat{s}) = \alpha_0/(1+\hat{s}/(2~\mathrm{TeV})^2)^2$, which turns off
the couplings at very large $\hat{s}$ where the couplings would violate unitarity.
In order to further simplify the coupling parameter space, the equal coupling
scheme is used,  $\Delta\kappa\equiv{\kappa_\gamma}-1 = {\kappa_{Z}}-1$ and 
$\lambda\equiv{\lambda_{Z}}={\lambda_{\gamma}}$ with ${g_1^Z} = 0$. Other
simplifications of the parameter space have also been studied.

\Dzero sets limits on the anomalous coupling constants using the 
$ll'\nu\nu$ sample by fitting
the leading lepton $\pt$ spectrum  as shown in Figure \ref{fig:TGC_fit}a). The figure
shows the effect on the shape of the spectrum due to anomalous couplings
near the current bounds.
CDF performs a similar fit to the $\pt$ spectrum of the leptonically decaying
$W$ in the $l\nu jj$ final state (Figure \ref{fig:TGC_fit}b). The results 
of these fits are $-0.32<{\Delta\kappa}<0.45$ and $-0.29<{\lambda}<0.30$ (D\O\ $\! \! \!$)
and  $-0.51 < {\Delta \kappa} < 0.44$ and $ -0.28 < {\lambda} < 0.28$ (CDF).

\section{Search for $WZ$ production}
\label{sec:WZ}

Both CDF and \Dzero search for $WZ$ in the $lll'\nu$ final state
$(l,l'=e$ or $\mu$) which has a combined branching fraction of $1.8\%$,
including contributions from $\tau\rightarrow l\nu\nu$ where $l=e$ or $\mu$.
The small standard model prediction for the $WZ$ cross-section makes this a 
very low rate signal, so both experiments have optimized their lepton selection 
criteria to maximize efficiency and acceptance. 

\begin{figure}[htb]
\includegraphics[clip=true,width=0.5\textwidth]{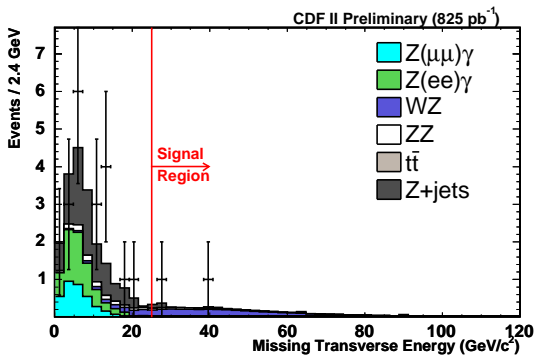}
\caption{The $\MET$ distribution for $WZ\rightarrow lll'\nu$ candidate events at CDF showing
the power of the $\MET$ requirement for suppressing backgrounds.}
\label{fig:CDF_WZ_MET}
\end{figure}

The dominant backgrounds in these searches are $Z\rightarrow ll$ with a jet or photon
misidentified as a lepton and $ZZ\rightarrow lll'l'$  where one lepton is not
reconstructed.
As shown in Figure \ref{fig:CDF_WZ_MET}, these backgrounds are both strongly 
suppressed by requiring the event to have $\MET>$  25 GeV (CDF), 20 GeV (D\O\ $\! \! \!$).

\begin{figure*}[!thb]
\includegraphics[width=0.48\textwidth]{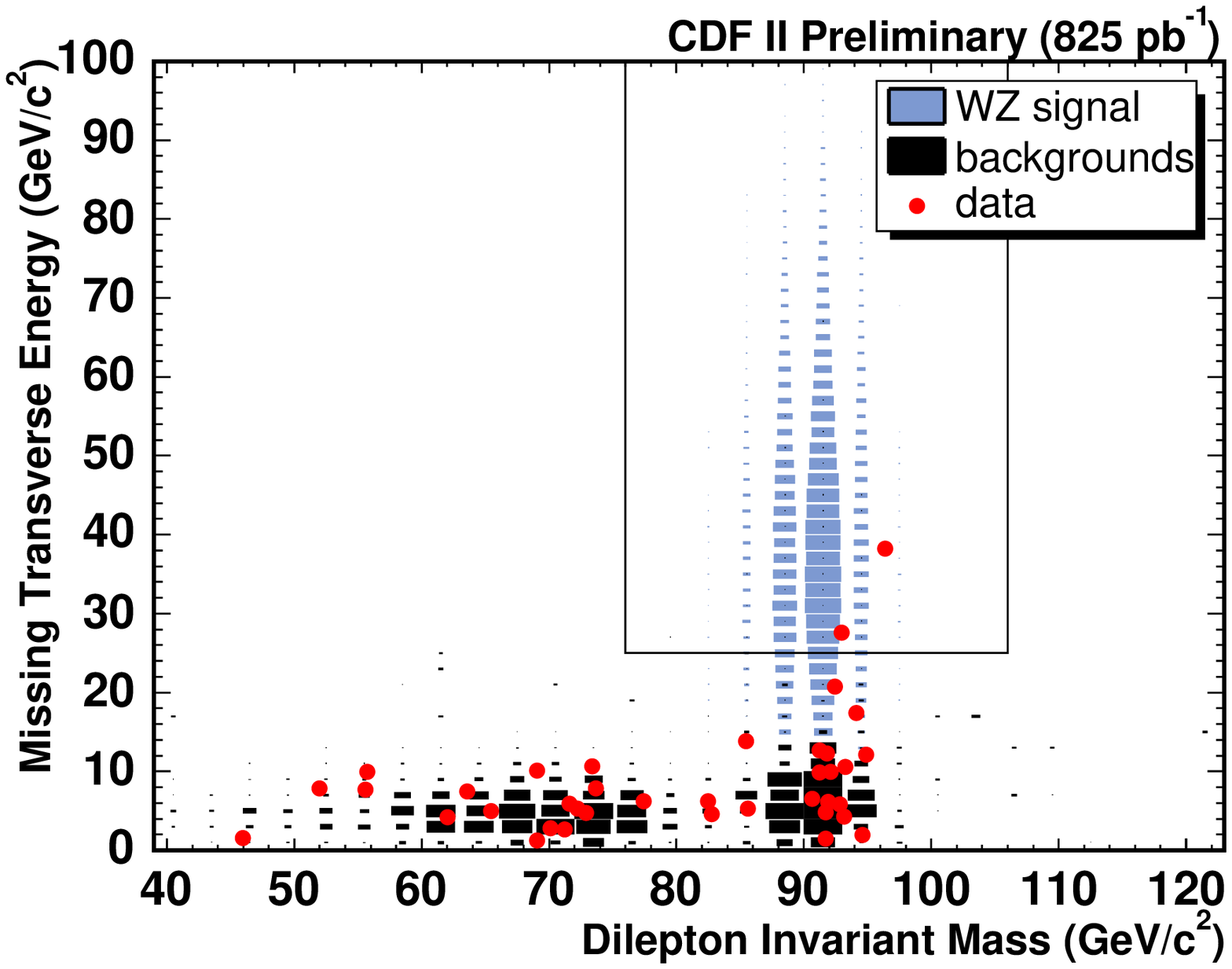}
\includegraphics[width=0.51\textwidth]{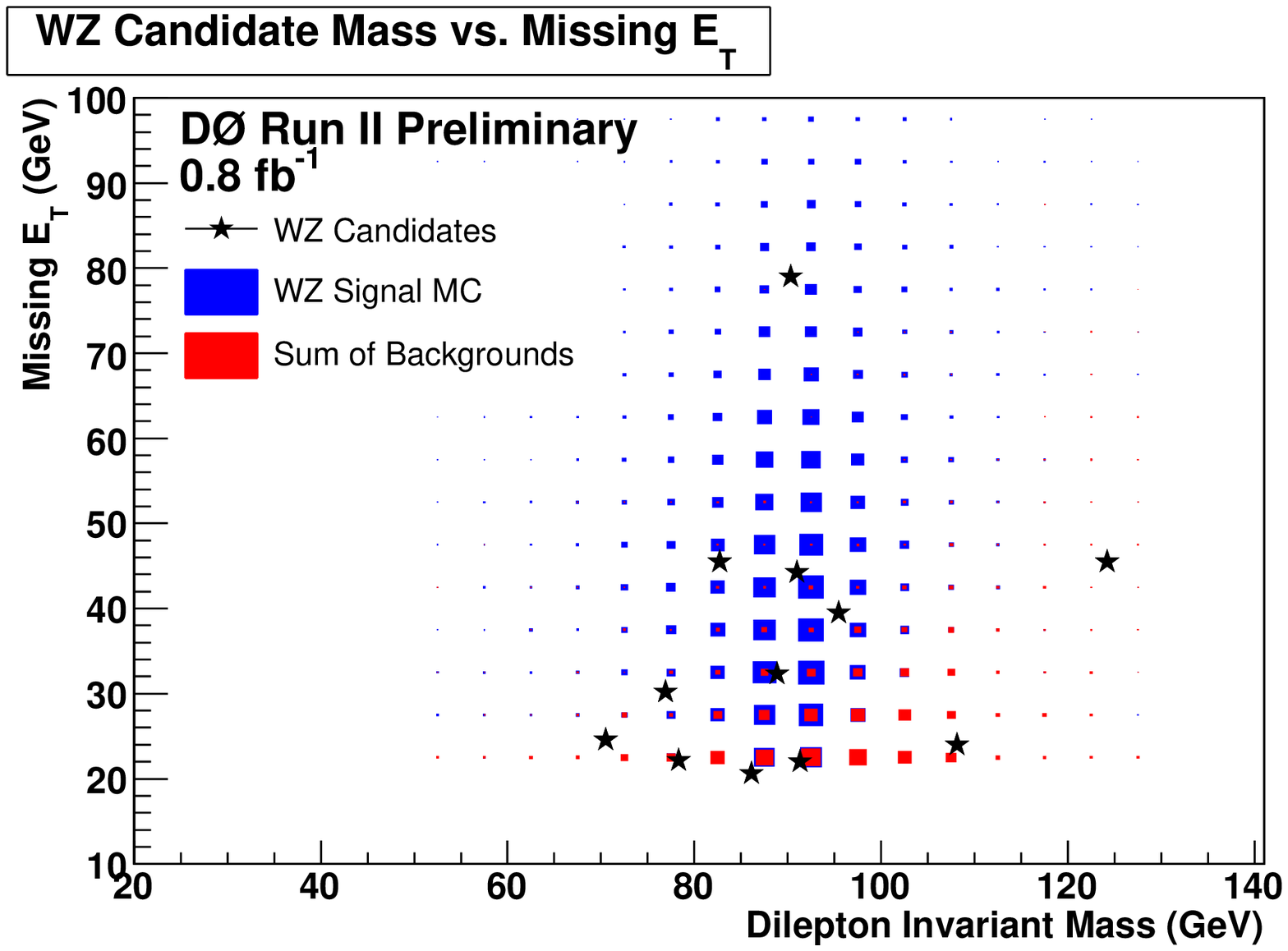}
\caption{The distributions of the observed events in the dilepton invariant mass versus $\MET$ plane
for (left) CDF and (right) \Dzero.}
\label{fig:WZ_2dplots}
\end{figure*}

CDF observes 2 events expecting 3.72 $\pm$ 0.02(stat.) $\pm$ 0.15(syst.) $WZ$
signal events and 0.92 $\pm$ 0.07(stat.) ${}^{+0.16}_{-0.10}$(syst.) background events 
using $825~\rm{pb}^{-1}$ of data. Based
on this CDF sets an upper limit of $\sigma(WZ) < 6.3~\rm{pb}$ at 95\% CL. \Dzero
observes 12 events with an expectation of $7.5\pm1.2$ signal and $3.6\pm0.2$ background 
events using $800~\rm{pb}^{-1}$ of data. This constitutes 3.3$\sigma$ evidence 
for $WZ$ production and corresponds to a cross-section of $\sigma(WZ) = 4.0^{+1.9}_{-1.5}~\rm{pb}$,
consistent with the standard model prediction. The distributions
of the observed events in the dilepton invariant mass versus $\MET$ plane
show that signal events cluster near the $Z$ mass and values of $\MET$ 
consistent with expectation (Figure \ref{fig:WZ_2dplots}).

\section{Summary}

With the increasingly large accumulated datasets at the Tevatron, CDF and \Dzero
are probing electroweak diboson production with new sensitivity. Cross-section
and triple gauge coupling measurements for the $WW$ final state are becoming
increasingly precise, while the first evidence for $WZ$ production has just been
seen.

\end{document}